\documentclass[twocolumn,twoside,slac_two]{revtex4}
\usepackage{graphicx}
\usepackage{fancyhdr}
\usepackage{hyperref}
\hypersetup{
    colorlinks=true,
    urlcolor=magenta,
}
\begin{document}

\title{The EUSO$@$TurLab Project}

%

\author{
  H. Miyamoto, M. Bertaina, G. Cotto, R. Forza, M. Manfrin, M. Mignone,
  G. Suino, A. Youssef
}
\affiliation{
  Universit$\grave{a}$ di Torino/INFN Torino, 
  via Pietro Giuria 1, 10125 Torino, Italy
}
\author{
  R. Caruso, G. Contino
}
\affiliation{
  Universit$\grave{a}$ di Catania/INFN Catania, 
  via Santa Sofia 64, 95123 Catania, Italy
}
\author{
  S. Bacholle, P. Gorodetzky, A. Jung, E. Parizot, G. Prev$\hat{o}$t
}
\affiliation{
  Laboratoire APC
  10, rue Alice Domon et L$\acute{e}onie$ Duquet, 75013 Paris, France
}
\author{
  P. Barrillon, S. Dagoret-Campagne, J. Rabanal Reina
}
\affiliation{
  LAL/IN2P3/CNRS/Universit$\acute{e}$ Paris-Sud\\
  Centre Scientifique d'Orsay, B$\hat{a}$timent 200 - BP 34
  91898 Orsay cedex, France
}
\author{
  S. Blin
}
\affiliation{
  OMEGA/CNRS/IN2P3,
  Ecole Polytechnique
  91128 Palaiseau Cedex, France
}
\author{
  \vspace{-0.3cm}
  for the JEM-EUSO Collaboration
}

\begin{abstract}
The TurLab facility is a laboratory, equipped with a 5 m diameter and 1 m depth rotating tank,
located in the Physics Department of the University of Turin.
The tank has been built mainly to study problems where system rotation plays a key role in the fluid behaviour such as in atmospheric and oceanic flows at different scales.
The tank can be filled with different fluids of variable density, which enables studies in layered conditions such as sea waves.\\
The tank can be also used to simulate the terrestrial surface with the optical characteristics of different environments such as snow, grass, ocean, land with soil, stones etc., fogs and clouds.
As it is located in an extremely dark place, the light intensity can be controlled artificially.
Such capabilities of the TurLab facility are applied to perform experiments related to the observation of Extreme Energy Cosmic Rays (EECRs) from space using the fluorescence technique, as in the case of the JEM-EUSO mission, where the diffuse night brightness and artificial light sources can vary significantly in time and space inside the Field of View (FoV) of the telescope.\\
Here we will report the currently ongoing activity at the TurLab facility in the framework of the JEM-EUSO mission (EUSO@TurLab).
\end{abstract} 
%
\maketitle
\thispagestyle{fancy}
%
%
%
\vspace{-0.5cm}
\section{JEM-EUSO and its pathfinders}
\vspace{-0.3cm}
JEM-EUSO~\cite{jemeuso} is the concept of a space-borne fluorescence telescope to be hosted on the International Space Station (ISS).
One of its main goals is a high statistical observation of EECRs with primary particle energies above 5$\times10^{19} eV$.
The telescope consists of three Fresnel lenses and a focal surface consisting of 0.3M pixels of UV sensitive detectors with a fast readout system, which enables a single photon counting measurement with a wide FoV of $\pm30^{\circ}$ with a spatial resolution of $500\times500$ $m^2$ on the ground, covering roughly 60 $\%$ of the entire surface of the Earth in its flight on the ISS orbit.
Looking towards the Earth from space, JEM-EUSO will reveal these particles as well as very high energy neutrinos by observing the fluorescence signal from the generated Extensive Air Showers (EAS) during their passage through the atmosphere.
It will also contribute to the investigation of atmospheric phenomena such as Transient Luminous Events (TLEs) and meteors.\\
The JEM-EUSO project is carried out by an international collaboration consisting of currently 88 institutions from 16 countries.
In parallel to the JEM-EUSO development, several pathfinders such as EUSO-Balloon~\cite{eusoballoon}, launched in August 2014, and a ground-based pathfinder TA-EUSO~\cite{taeuso}, currently in operation at the Black Rock Mesa in Utah, US have been developed.
Also, a space-borne pathfinder MINI-EUSO as well as an advanced balloon-borne pathfinder with NASA Super Pressure Balloon, EUSO-SPB, are currently being developed to be launched in 2017.\\
%
%
The ISS is flying at a speed of 7.5 km/s, which makes 15.5 orbits per day, and every 30 to 45 minutes it passes by the night region.
While the ISS flies on the orbit, it passes by many kinds of sceneries such as oceans, clouds, city light, airglows, forests, lightning, and so on from the altitude of 400 km.
\begin{figure*}
\centering
\includegraphics[width=0.68\hsize,height=3.3cm]{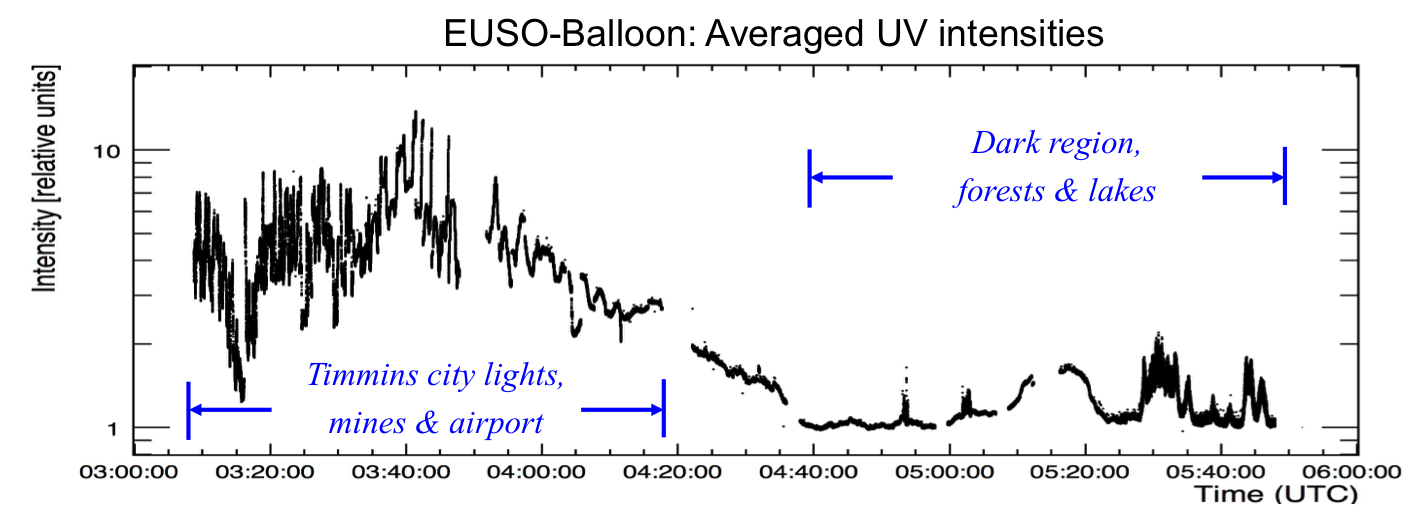}
\vspace{-0.5cm}
\caption{EUSO-Balloon Results on UV intensities.} 
\vspace{-0.2cm}
\label{EBresult}
\end{figure*}
\begin{figure*}[t]
\includegraphics[width=0.69\hsize,height=6.3cm]{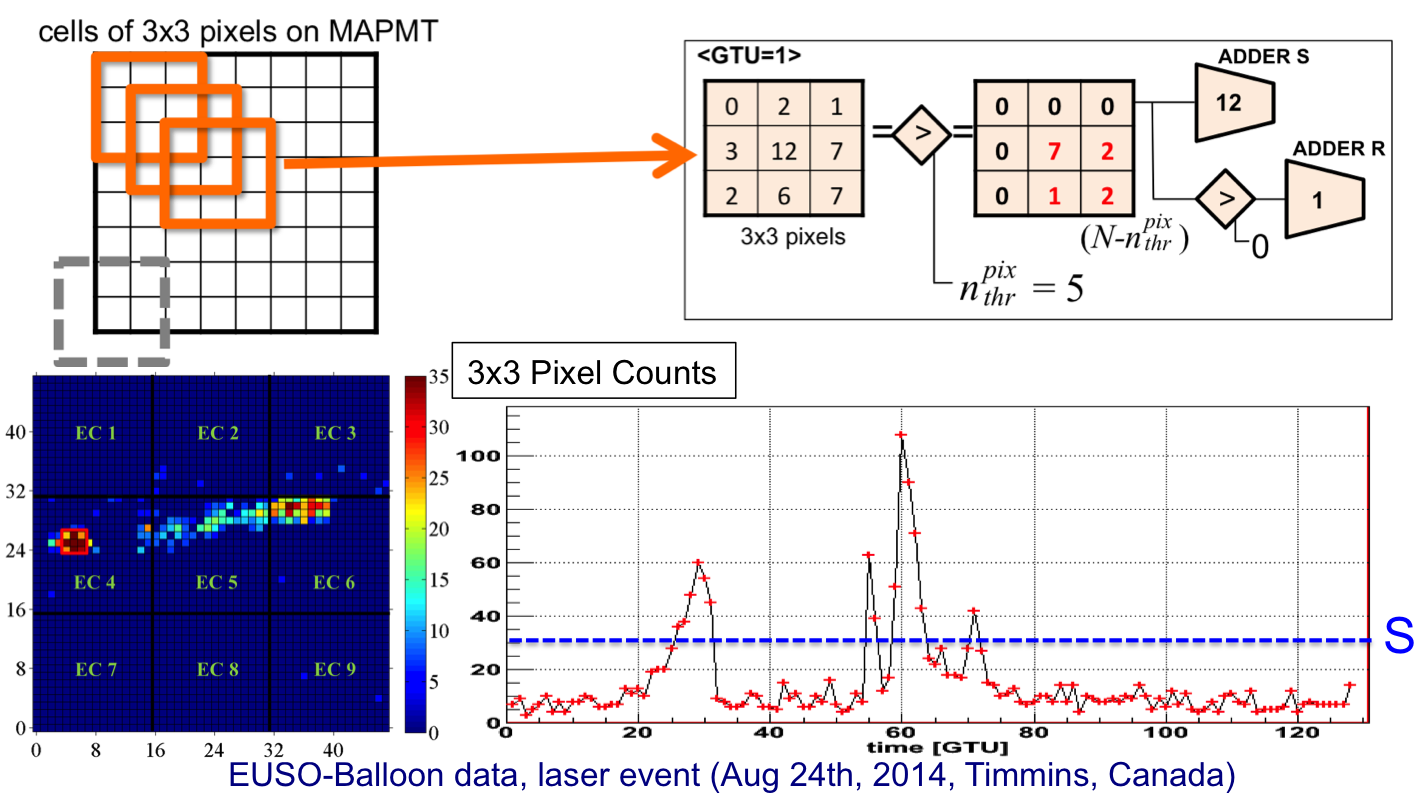}
\vspace{-0.3cm}
\caption{
  Top: conceptual figure of FLT.
  Bottom: integrated image (left) and 3$\times$3 pixel counts (right) during an EUSO-Balloon laser event.
}
\vspace{-0.3cm}
\label{EUSOflt}
\end{figure*}
The Fig.~\ref{EBresult} shows the UV intensity measured during the five hours of an EUSO-Balloon flight in August 2014.
The balloon was flying from left to right, from Timmins airport to the dark region where mostly lakes and forests are located.
The plot shows the averaged intensity in reference area in logarithmic scale as a function of time (UTC).
The high intensities at the beginning of the flight (left part in the plot) appeared when the balloon was flying above cities like Timmins and neighbourhoods, mines and airports,
where there are full of artificial lights.
One may see the UV background level is increasing by a factor of 10 comparing to other areas mainly covered by forests and lakes.
\vspace{-0.3cm}
\subsection*{JEM-EUSO Focal Surface and Data Acquisition Chain (DAQ)}
\vspace{-0.3cm}
JEM-EUSO focal surface is consisting of Hamamatsu 64-ch Multi-Anode PhotoMultiplier Tubes (MAPMTs).
One Elementary Cell unit (EC\_unit) consists of 4 PMTs, and 9 EC\_units form a Photo-Detector Module (PDM).
In total, the entire focal surface consists of 137 PDMs with about 5,000 PMTs with 0.3M pixels. 
The output of PMTs are readout by an Elementary Cell ASIC board (EC\_ASIC) which consists of 6 SPACIROC ASICs~\cite{spaciroc}.
A PDM board is an interface board between a PDM and the latter part of data processing system, which sends slow control commands and processes First Level Triggers (FLTs).
%
%
%
\vspace{-0.6cm}
\subsection*{First Level Trigger (FLT)}
\vspace{-0.4cm}
JEM-EUSO will have to deal with a huge amount of data.
Therefore, it is essential to develop an efficient FLT which detects UHECRs in a continuously varying background.
The Fig.~\ref{EUSOflt} shows a conceptual figure of EUSO FLT (top)
with an example of an EUSO-Balloon laser event~\cite{laser} (bottom); 
i.e., an integrated image during 320 $\mu$s (left) and 3$\times$3 pixel counts as a function of frame (Gate Time Unit, GTU=2.5$\mu$s, right).
The FLT is a persistence trigger which scans counts in every 3$\times$3 pixel-boxes in a PMT, checking the excess against the pixel threshold which is automatically adjusted in every 320 $\mu$s.
FLT sets a proper threshold automatically as indicated by the blue dotted line in the  plot
to avoid triggering on the background, but only on the EAS-like signals.
%
\begin{figure*}
\includegraphics[width=0.6\hsize,height=6.8cm]{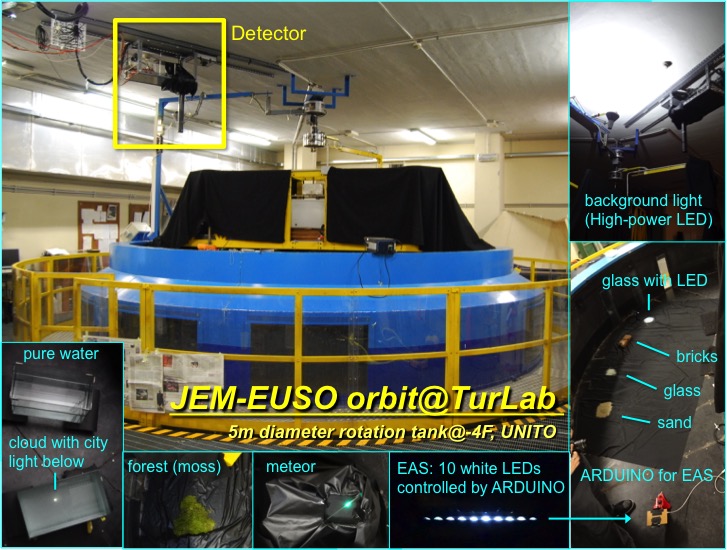}
\vspace{-0.3cm}
\caption{
  The TurLab rotating tank. 
  The black tube on the ceiling shows the collimator of the experimental setup used to mimic the JEM-EUSO telescope.
  Light sources and materials used to mimic other phenomena are also shown.
}
\vspace{-0.3cm}
\label{vasca}
\end{figure*}
\begin{figure*}
\includegraphics[width=0.9\hsize]{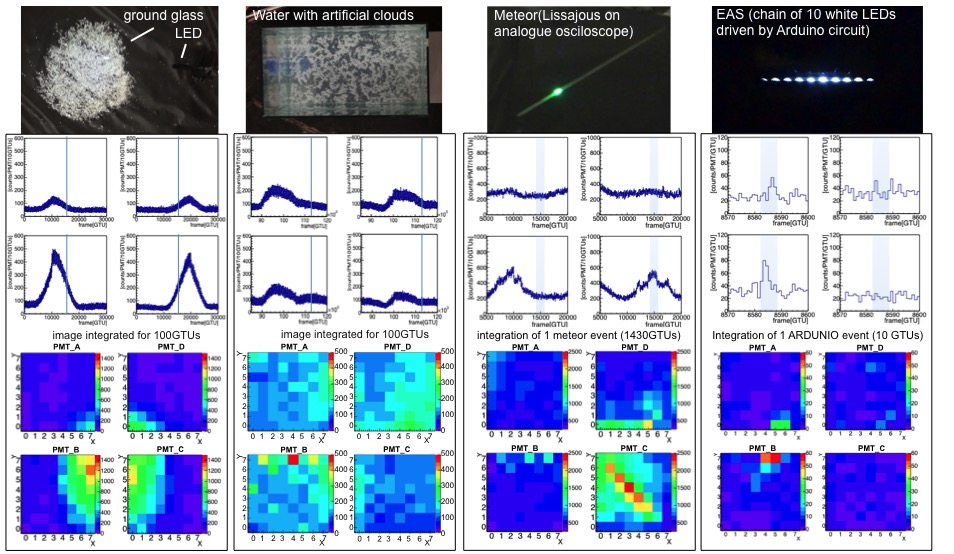}
\vspace{-0.3cm}
\caption{
Examples of UV images obtained by the EC\_unit (2$\times$2 MAPMTs) camera passing by the various materials during the full tank rotation ($\sim$9 min).
Starting from the left, ground glass illuminated by LED (city light), cloud, meteor and EAS like events are mimicked as shown in the photos on the top.
Middle plots show the summed counts per each PMTs per appropriate frames as a function of frame (GTU), while bottom plots show the images obtained during each events integrated during the frames indicated by blue vertical lines in middle plots.
}
\vspace{-0.4cm}
\label{9minImages}
\end{figure*}
%
%
%
%
\begin{figure*}[t]
\includegraphics[width=0.76\hsize,height=8.3cm]{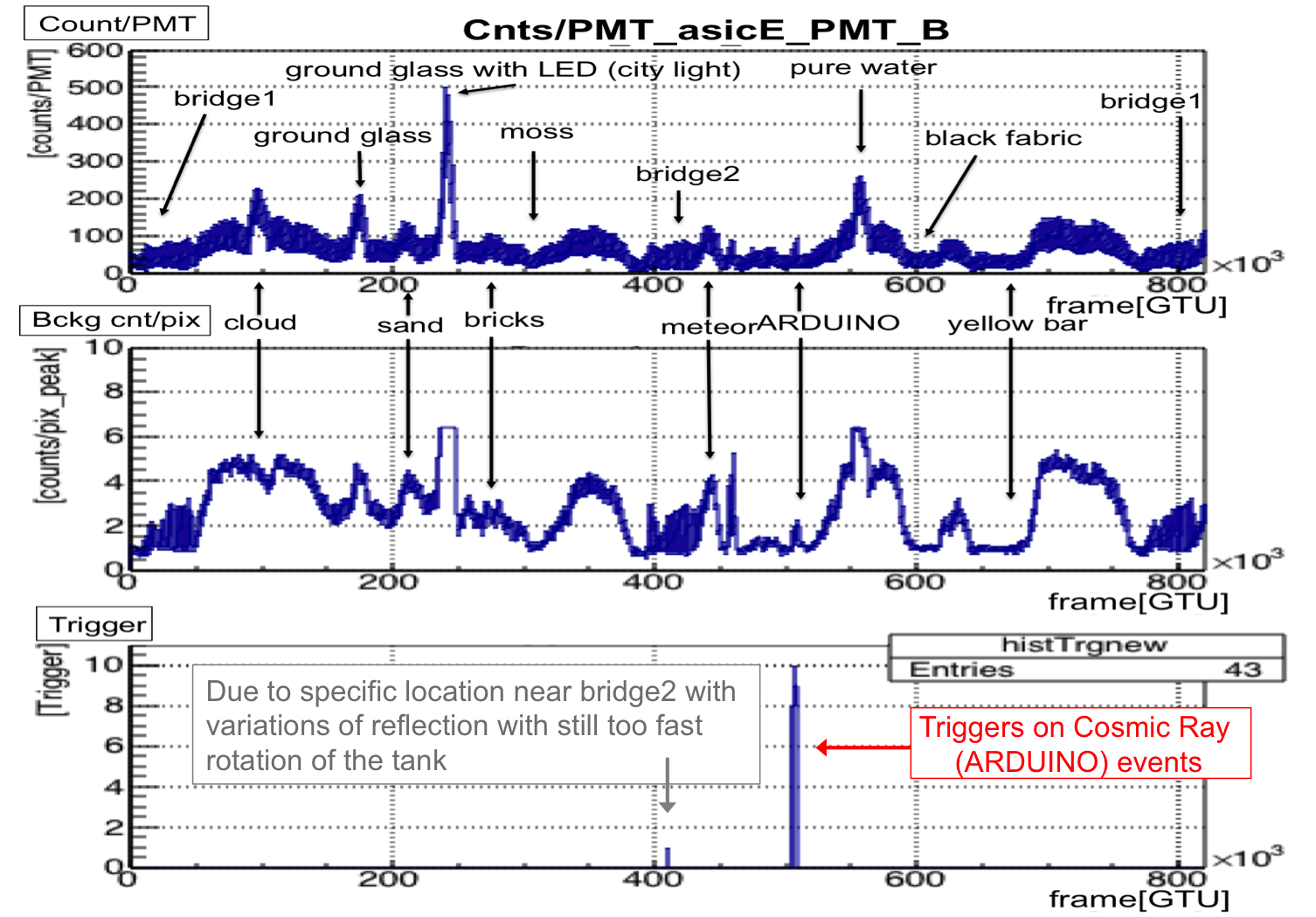}
\vspace{-0.3cm}
\caption{
The top panel shows the raw data (counts/PMT), while the middle shows the corresponding background level which is used to set the FLT threshold and the bottom shows when FLTs were issued, as a function of frame (GTU) respectively.
We obtained 100\% of the trigger efficiency against the cosmic-ray-like events (LED chain driven by ARDUINO circuit).
}
\vspace{-0.5cm}
\label{FLToffline}
\end{figure*}
\vspace{-0.5cm}
\section{The TurLab@Physics Department - Torino University}
\vspace{-0.3cm}
To test our electronics, as well as study and develop the FLT in such a dynamic condition, we got an idea to use a big rotation tank of TurLab~\cite{turlab}, which is located in the fourth basement of the physics department building of University of Turin.\\
TurLab is a laboratory for geo-fluid-dynamics studies, where rotation is a key parameter such as Coriolis force and Rossby Number.
With using inks or particles, based on the fluid-dynamics, key phenomena such as planetary atmospheric and fluid instabilities can be reproduced in the TurLab water tank.
The tank has 5 m diameter with a capability of the rotation at a speed of 3 s to 20 min per rotation.
Also, as it is located in a very dark environment, the light intensity can be controlled artificially.
%
\vspace{-0.5cm}
\section{The EUSO@TurLab Project}
\vspace{-0.3cm}
By means of the rotating tank with the capabilities mentioned above, we have been testing EUSO electronics such as its basic performance as well as the FLT for cosmic rays in the various background and sceneries which are transiting from one to another.\\
When we set an EUSO EC\_unit camera, a 2$\times$2 array of 64-ch MAPMTs on the ceiling at the height of 2 meters above the TurLab tank, we know that a pixel watches a FoV of 5$\times5$ $mm^2$ from the past measurements.
This FoV corresponds to a solid angle of 6.25$\times10^{-6} sr$ which is comparable to the one for JEM-EUSO.
Therefore, considering the altitude and the speed of ISS, we learned that we can reproduce the JEM-EUSO observation in TurLab.
The Fig.~\ref{vasca} shows the TurLab tank, light sources and materials used to mimic the various kinds of phenomena and albedos that JEM-EUSO will encounter. 
We use an EUSO EC\_unit with a lens as a camera, readout by the EUSO front-end electronics with a test board, observing several materials such as meteor, cosmic ray, cloud, ground glass, passing from one to another in the FoV in a constant background light produced by a high power LED lamp diffused on the ceiling.
High voltages, DC power supplies, function generators and monitoring oscilloscopes are on the desk by the side of the tank with a PC with LabView interface for the test board to send commands for slow control and data acquisition, and  ROOT programs for monitoring and analysis.
Fig.~\ref{9minImages} 
shows the examples of UV images obtained by the EC\_unit camera during the full tank rotation. 
%
%
%
\vspace{-0.7cm}
\subsection*{FLT analysis (Offline)}
\vspace{-0.3cm}
The top panel of Fig.~\ref{FLToffline} shows the result of UV intensities in a full rotation of the tank with various materials.
The plot shows the summed counts of a PMT ($=$64 pixels) as a function of time in a unit of frame.
%
Afterward, we analysed the data and processed the FLT offline.
The middle panel shows the averaged counts per pixel which is used to set the FLT threshold, while the bottom shows when FLTs were issued based on signals in that PMT, as a function of frame (GTU) respectively.
Almost all triggers coincide with passing over the ARDUINO driven LED chain, which mimics cosmic-ray like events, as it should be; the one that is not is due to a specific location near one of the two bridges crossing the tank where the variations of light reflection were still too fast to be compensated by the slower rotation of the tank.
%
\section{Summary}
%
JEM-EUSO and its pathfinders will observe various cosmic/atmospheric phenomena in UV from the space/atmosphere. 
TurLab is a unique facility in Turin with interdisciplinary experts (waves, geophysics, atmospheric science, meteors, astroparticle/cosmic ray physics, etc.), capable of providing an ideal condition to test EUSO electronics in a controlled environment against various sceneries which the telescope will encounter.
The TurLab system has been used to verify/implement FLT in use of EUSO-SPB. Analysis and development of FLT for Mini-EUSO are currently ongoing.\\
%
%
\begin{acknowledgments}
This work has been partially funded by the Italian Ministry of Foreign Affairs and International Cooperation, the European High-Performance Infrastructures in Turbulence (EuHIT).
\end{acknowledgments}

\end{document}